\documentstyle[12pt]{article}
\newcommand{\be}{\begin{equation}}
\newcommand{\ee}{\end{equation}}

\begin{document}

\title{{\LARGE Polarons in Wurtzite Nitride Semiconductors}}

\bigskip

\author{{\bf M. E. Mora-Ramos\thanks{Present Address: Facultad de Ciencias. 
Universidad Aut\'onoma del  Estado de Morelos . Ave. Universidad 1001. CP. 62210. 
Cuernavaca, Morelos, M\'exico}, F. J. Rodr\'{\i}guez, and L. Quiroga}\\
Departamento de F\'{\i}sica. Universidad de Los Andes\\
A.A. 4976, Santaf\'e de Bogot\'a, Colombia}

\date{}
\maketitle
\begin{center}
(Received Sept. 3, 1998 by C.E.T. Gon\c{c}alves Da Silva)
\end{center}

\baselineskip 7mm

\begin{abstract}
Polaron binding energy and effective mass are calculated for
semiconductors with wurtzite crystalline structure from the first
order electron-phonon corrections to the self-energy. A recently introduced
Fr\"ohlich-like electron-phonon interaction   
Hamiltonian which accounts for the LO and TO polarizations mixing due to
the anisotropy is used in the calculation. The polaronic damping rates are 
evaluated for finite temperature. Numerical 
results are  reported for 
$GaN$. It is shown that the electron-phonon coupling is strong enough 
to justify the necessity of the inclusion of second-order corrections.
\end{abstract}

\vspace{2.5cm}


\newpage


Electronics and optoelectronics based on III-V wide band gap nitride 
semiconducting systems have emerged as a matter of particular interest in recent 
years. An increasing number of papers devoted to the study of these materials 
and their applications can be found in the literature (see, for example, Refs. 
$[1-4]$).

Nitrides have the wurtzite as their natural crystalline structure, and exhibit 
highly unusual properties among the III-V group of semiconductors, more 
resembling, in some aspects,  the II-VI compounds. It is possible to mention, for 
instance, the large magnitudes of the spontaneous and piezoelectric polarizations $[5]$.

An aspect which is only beginning to be investigated in crystals with such a 
structure is the carrier-polar-optical-phonon interaction mechanism. A 
Fr\"ohlich-type electron-phonon Hamiltonian was recently put forward $[6]$ within 
the non-retarded Born-Huang dielectric continuum approach. In that work only the three
optical-phonon 
branches which are infrared-active in wurtzites are considered. Probably, the most 
important result arising from that work is the mixing of the longitudinal and 
transverse optical
 modes due to the crystalline anisotropy. As a consequence, both kind of modes 
do contribute to the carrier scattering, although LO-like contributions should be dominant. 
A similar behaviour is 
well known in semiconducting heterostructures, where the presence of interfaces 
breaks the isotropy in one or more spatial directions, 
leading to the mixing of the longitudinal and transverse oscillation polarizations 
$[7,8]$ , and all of them have to be included in the study of the electron-phonon 
interaction because even the TO-type modes can provide a non-negligible 
contribution $[9]$
 to the scattering rates and polaron properties.

Furthermore, the existence of a spatial direction with lesser symmetry also gives 
rise to phonon dispersion. To illustrate, let us mention the application of the 
macroscopic dielectric continuum model to the study of the long-wavelength 
optical oscillations in dielectric or semiconducting heterolayers $[10]$. In that 
case, even when no phonon dispersion is considered for the bulk constituent 
materials, a dispersion law $\omega(q)$ appears for the so-called interface 
modes. For wurzite
 III-V nitrides, phonon dispersion effects should be expected. As already discussed 
in Ref. $[6]$, the 
dependence of the oscillation frequencies upon the phonon wavevector ${\bf q}$ is 
not a function  of the wavevector magnitude but of the angle between ${\bf q}$ and 
the $c$ axis of the structure (referred to from now on as z direction),
the direction of less symmetry in the unit 
cell.

The electron-phonon scattering rates $[6]$ 
are sufficiently large to recognize a rather strong 
interaction, and to motivate the study of the polaronic corrections to the 
energy and the effective mass of slow electrons in the conduction band of wurtzite 
nitride materials.

The polaron is a quasiparticle state consisting of an electron and its surrounding 
phonon cloud. When electrons move through a polar material, they polarize the 
surrounding medium and couple to the self-induced polarization field $[11]$. 
This coupling slows down the electron as if it
acquired an additional mass, and -at the same time- behaves as a potential well, 
decreasing its zero momentum energy, thus tending to localize the electron (which 
really happens when the coupling is strong enough).
Even for weakly polar materials, these 
corrections may be 
quite important and will necessarily have to be taken into account, for instance, 
in the study of optical and transport properties. The aim of this 
work is -then- to calculate the polaronic corrections in wurtzite
nitride  semiconducting 
materials according to the dielectric long-wavelength continuum model. In particular, 
numerical results will be presented for $GaN$.


The electron-phonon interaction Hamiltonian corresponding to 
the infrared-active phonon modes 
of wurtzite structures can be written in general form as the sum of two 
contributions [6]: one of them corresponds to 
electron-LO-like-phonon interaction and the second one stands for the 
electron-transverse-optical(TO)-like-phonon interaction. The longitudinal and 
transverse electron-phonon vertices are given by$[6]$

\begin{eqnarray}
\vert M^{L}_{{\bf q}}\vert^2=\frac{2\pi 
e^2\hbar}{Vq^2\Omega_{L}}\left[\frac{\sin^2\theta}{(1/\epsilon^{*}_{\perp})
\omega^{2}_{\perp 
L}}+\frac{\cos^{2}\theta}{(1/\epsilon^{*}_{z})\omega^{2}_{zL}}\right]^{-1}, 
\nonumber\\
\vert M^{T}_{{\bf q}}\vert^2=\frac{2\pi 
e^2\hbar}{Vq^2\Omega_{T}}\frac{(\omega^{2}_{\perp}-\omega^{2}_{z})^2 
\sin^{2}\theta 
\cos^{2}\theta}{(\epsilon^{0}_{\perp}-\epsilon^{\infty}_{\perp})
\omega^{2}_{\perp}
\cos^{2}\theta+(\epsilon^{0}_{z}-\epsilon^{\infty}_{z})\omega^{2}_{z}\sin^{2}\theta}.
\end{eqnarray}

Here, $1/\epsilon^{*}_{\perp}=1/\epsilon^{\infty}_{\perp}-1/\epsilon^{0}_{\perp}$ 
and $1/\epsilon^{*}_{z}=1/\epsilon^{\infty}_{z}-1/\epsilon^{0}_{z}$; 
$\epsilon^{\infty}_{\perp}(\epsilon^{\infty}_{z})$ being the 
high-frequency  dielectric  constant perpendicular to (along) the $z$ axis, and 
$\epsilon^{0}_{\perp}=\epsilon^{\infty}_{\perp}\omega^{2}_{\perp L}/\omega^{2}_{\perp}$; 
$\epsilon^{0}_{z}=\epsilon^{\infty}_{z}\omega^{2}_{zL}/\omega^{2}_{z}$ are the 
static dielectric constants. $\omega_{zL}$ and $\omega_{\perp L}$ are the 
LO-phonon frequencies along and perpendicular to the $z$ axis respectively, and 
$\omega_{z}$ and $\omega_{\perp}$ are the corresponding lattice dispersion frequencies. 
For simplicity, $\epsilon^{\infty}_{z}$ and $\epsilon^{\infty}_{\perp}$ are 
assumed to be identical $[6]$.

The characteristic frequencies $\Omega_{L}$ and $\Omega_{T}$ are also functions of 
$\theta$, the angle between the wavevector ${\bf q}$ and the $z$ 
direction. They are given by$[6]$

\begin{eqnarray}
\Omega^{2}_{L}=\omega^{2}_{zL}\cos^{2}\theta+\omega^{2}_{\perp L}\sin^{2}\theta,\nonumber\\
\Omega^{2}_{T}=\omega^{2}_{z}\sin^{2}\theta+\omega^{2}_{\perp}\cos^{2}\theta,
\end{eqnarray}

Unlike isotropic materials, the TO-like vertex $M^{T}_{{\bf q}}$ is, in general, 
different from zero due to the polarization mixing. A pure transversal mode is 
obtained for $\theta=\frac{\pi}{2}$.

The electronic states are described, as usual, within the effective mass 
approximation: The electron wavevector ${\bf k}$ is the quantum number which 
labels the state with energy $E=\hbar^2k^2/2m^{*}$ in the conduction band, which 
is assumed to be spherically symmetric with an effective mass $m^{*}$. 

In order to illustrate the results of our calculations we choose $GaN$ with 
frequencies and dielectric constants: 
$\omega_{zL}=$735$\;cm^{-1}$, $\omega_{\perp L}=$743$\;cm^{-1}$ $[12]$, 
$\epsilon^{\infty}_{z}=$5.29, $\epsilon^{0}_{z}=$10.01 and 
$\epsilon^{0}_{\perp}=$9.28. The conduction band effective 
mass is taken to be $m^{*}=$0.20$m_{0}$ $[13]$, $m_{0}$ being the electron bare 
mass.

To deal with the entire range of $\theta$, a Fr\"ohlich-like coupling constant 
$\alpha_{L}$ is introduced through
$\alpha_{L}^{2}=m^{*}e^{4}/(2\epsilon^{*}_{\perp}\epsilon^{*}_{z}\hbar^{3}\omega_{
L})$. $\omega_{L}$ would be an "effective longitudinal frequency". In our case, we 
choose the value $\omega_{L}=\Omega_{L}(\theta=\pi/4)$ to treat contributions 
coming from 
$\omega_{\perp L}$ and $\omega_{zL}$ on equal footing. For $GaN$, 
$\hbar\omega_{L}=91.64\;meV$ and $\alpha_{L}=0.39$. An "effective longitudinal 
polaron radius"  $\rho_{L}$ introduced through  
$\rho^{2}_{L}=\hbar/2m^{*}\omega_{L}$ has the value 14.4
1\AA$\;\;$ for GaN.

To describe the electron-TO-like-phonon interaction in terms of certain 
dimensionless coupling parameter, let us introduce another Fr\"ohlich-like 
constant $\alpha_{T}$, so that 
$\alpha_{T}^{2}=m^{*}e^{4}/(2\epsilon^{*}_{\perp}\epsilon^{*}_{z}\hbar^{3}\omega_{T})$.
Under the same argument used to introduce the frequency $\omega_{L}$, an 
"effective transversal frequency" $\omega_{T}$ is chosen to be 
$\omega_{T}=\Omega_{T}(\theta=\pi/4)$.

For $GaN$ we get $\hbar\omega_{T}=67.86\;\;meV$, leading to the values 
$\alpha_{T}=0.45$, and $\rho_{T}=16.75$\AA,  for the coupling constant and the 
"effective transversal polaron radius", respectively.

Now, the longitudinal and transverse electron-phonon vertices given in  Eqn. (1), 
modify in the following way: 
the factor $2\pi e^{2}\hbar$ is changed to 
$4\pi\hbar^{2}\alpha_{L}((1/\epsilon^{*}_{z})(1/\epsilon^{*}_{\perp}))^{-1/2}\rho_{L
}\omega_{L}$ and to 
$4\pi\hbar^{2}\alpha_{T}((1/\epsilon^{*}_{z})(1/\epsilon^{*}_{\perp}))^{-1/2}\rho_{T}\omega_{T}$ for the longitudinal and the transversal case, respectively.

This way of choosing $\omega_{L}$, $\omega_{T}$, $\alpha_{L}$, and $\alpha_{T}$, 
to represent the electron-phonon interaction for the wurtzite is, of course,  not 
unique. It is possible to follow different criteria; but what remains clear is 
that the final numerical results will be the same. Indeed, it makes no sense to 
compare the strengths 
of the electron-LO-like-phonon interaction and of the electron-TO-like-phonon 
interaction through the values of $\alpha_{L}$ and $\alpha_{T}$. However, both 
$\alpha_{L}$ and $\alpha_{T}$ can be  representative values if comparisons between 
the strength of
 the corresponding interactions are made between different wurtzite materials.

Calculating the first order correction to the electron self-energy, the real and 
imaginary parts will be given by $[14]$

\begin{eqnarray*}
Re \left[ \Sigma^{(1)}_{S}(k,\gamma,\eta) \right]\!=\!-\frac{m^{*}}{\pi^{2}} 
\!\!\left(\epsilon^{*}_{z}\epsilon^{*}_{\perp}\right)^{1/2}
\!\!\!\!\!\!\!\!\alpha_{S}\rho_{S}\omega_{S}\!\!\int^{\infty}_{0}\!\!\!\!dq\!\!
\int^{2\pi}_{0}\!\!\!\!d\varphi\!\! \int^{1}_{-1}\!\!dx 
\frac{f_{S}(x)}{\Omega_{S}(x)}\!\!\! \left[ \frac{N_{S}(x)+f_{{\bf k}-{\bf q}}}
{q^{2}-2qk\cos\nu-\frac{2m^{*}\Omega_{S}(x)}{\hbar}} \right. \\
+ \left. \frac{N_{S}(x)+1-f_{{\bf k}-{\bf 
q}}}{q^{2}-2qk\cos\nu+\frac{2m^{*}\Omega_{S}(x)}{\hbar}} \right], 
\end{eqnarray*}

\begin{eqnarray}
Im \left[ \Sigma^{(1)}_{S}(k,\gamma,\eta) \right]\!=\! -\frac{m^{*}}{\pi}\!\! 
\left(\epsilon^{*}_{z}\epsilon^{*}_{\perp} \right)^{1/2}\!\!\! 
\alpha_{S} \rho_{S} \omega_{S} \int^{\infty}_{0}\!\!dq\!\! \int^{2\pi}_{0}\!\! 
d\varphi \!\! \int^{1}_{
-1} \!\! dx \frac{f_{S}(x)}{\Omega_{S}(x)} \left[ \left( N_{S}(x)+f_{{\bf k}-{\bf 
q}} \right) \times \right. \nonumber \\
\left. \delta \left( q^{2}-2qk\cos\nu- \frac{2m^{*}\Omega_{S}(x)}{\hbar} \right) + 
\left( N_{S}(x) +1- f_{{\bf k}-{\bf q}} \right) \delta \left( q^{2}- 2qk\cos\nu + 
\frac{2m^{*}\Omega_{S}(x)}{\hbar} \right) \right];
\end{eqnarray}

with $x=\cos\theta$,
$\cos\nu=\cos(\varphi-\gamma)\sin\eta (1-x^{2})^{1/2}+x\cos\eta$, is the cosinus 
of the angle between the wavevectors ${\bf k}$ and ${\bf q}$, $\gamma$ is the 
polar angle associated to the electron wavevector ${\bf k}$, and $\eta$ is the 
angle between ${\bf k}$ and the $z$ direction.
The direction dependent characteristic phonon frequencies are given by

\be
\Omega^{2}_{S}(x)= \left\{\begin{array}{ll}
(\omega^{2}_{zL}-\omega^{2}_{\perp L})x^{2}+\omega^{2}_{\perp L},
&S=L; \\
\\
(\omega^{2}_{\perp}-\omega^{2}_{z})x^{2}+\omega^{2}_{z},
&S=T,
\end{array}
\right.
\ee

and

\be
f_{S}(x)=
\left\{\begin{array}{ll}
\left[\frac{1-x^{2}}{(1/\epsilon^{*}_{\perp})\omega^{2}_{\perp 
L}}+\frac{x^{2}}{(1/\epsilon^{*}_{z})\omega^{2}_{zL}}\right]^{-1},
&S=L; \\
\\
\frac{(\omega^{2}_{\perp}-\omega^{2}_{z})^2 
x^{2}(1-x^{2})}{(\epsilon^{0}_{\perp}-\epsilon^{\infty}_{\perp})\omega^{2}_{\perp}
x^{2}+(\epsilon^{0}_{z}-\epsilon^{\infty}_{z})\omega^{2}_{z}(1-x^{2})},
&S=T.
\end{array}
\right.
\ee
 
The polaron binding energy and effective mass are readily calculated from the real 
part of the electron-phonon retarded self-energy in the limits of one particle and 
zero temperature $[12]$. As usual, only the terms up to second order in $k$ 
are kept.
 Accordingly, the polaron binding energy will be given by,

\be
\varepsilon_{p}=-(\alpha_{L}\,\xi_{L}\,\hbar\,\omega_{L}+\alpha_{T}\,\xi_{T}\,\hbar\,\omega_{T}),
\ee

On the other hand, the polaron mass is given by

\be
m_{p}=\frac{1}{1-\mu_{p}(\eta)},
\ee

\be
\mu_{p}(\eta)=(\alpha_{L}\,\mu_{L1}+\alpha_{T}\,\mu_{T1})\,\cos^{2}\eta + 
(\alpha_{L}\,\mu_{L2}+\alpha_{T}\,\mu_{T2})\,\sin^{2}\eta.
\ee

Hence, the polaron effective mass in wurtzite crystals, with an 
electron-polar-optical-phonon interaction described by Eqn. (1) is  a function 
of the angle between the electron wavevector ${\bf k}$ and the $c$ axis of the 
crystal lattice. The expressions for the quantities $\xi_{S}$ and $\mu_{S}$ are the following $(S=L,T)$:

\be
\xi_{S}=\left(\frac{\epsilon^{*}_{z}\epsilon^{*}_{\perp}}
{\omega_{S}}
\right)^{1/2}\int^{1}_{0}\frac{f_{S}(x)}{\Omega^{3/2}_{S}(x)}dx;
\ee

\begin{eqnarray}
\mu_{S1}=\frac{1}{2}\left({\omega_{S}\epsilon^{*}_{z}\epsilon^{*}_{\perp}}
\right)^{1/2}\int^{1}_{0}\frac{x^{2}f_{S}(x)}{\Omega_{S}^{5/2}(x)}dx, \nonumber\\
\mu_{S2}=\frac{1}{4}\left({\omega_{S}\epsilon^{*}_{z}\epsilon^{*}_{\perp}}
\right)^{1/2}\int^{1}_{0}\frac{(1-x^{2})f_{S}(x)}{\Omega_{S}^{5/2}(x)}dx;
\end{eqnarray}

For $GaN$, the following numerical results were obtained: 
$\xi_{L}=9.86\times 10^{-1}$;
$\xi_{T}=3.79\times 10^{-3}$; $\mu_{L1}=1.68\times 10^{-1}$; $\mu_{L2}=1.61\times 
10^{-1}$; $\mu_{T1}=8.08\times 10^{-4}$, and $\mu_{T2}=5.46\times 10^{-4}$. Then, 
the numerical value of the polaron binding energy will be 
$\varepsilon_{p}=-(35.25+0.12)\;m
eV=-35.35\;meV$. Fig. 1 shows the relative magnitude of the 
effective mass first correction, $Mr=m_{p}/m^{*}-1$, versus $\eta$ for 
GaN. It  is seen that the average magnitude of the increasing of the electron 
effective mass
 is around $7\%$, and  the angular variation of this quantity is of approximately 
$4\%$, from $\theta=0$ to $\theta=\pi/2$.

The amount of the total corrections due to the electron-phonon interaction is, in 
fact, rather significative. Indeed, the relative weight of the contributions 
coming from the TO-like modes is very small: $0.34\%$ to the binding energy 
correction and $0.38
\%$ -in average- to the effective mass correction. As can be expected, the main 
contribution comes from the LO-like modes.

Anisotropy in the polaron effective mass is shown to be, in fact,
a very small effect. It is worth noting that in this work an 
isotropic spherical effective mass is assumed. However, in wurtzite structures
the electron effective mass should have
different values for $m^{*}_{z}$ and $m^{*}_{\perp}$. This fact could be
easily included in this formalism taking into account in Eq.(3) the corresponding electron wave function and 
energy dispersion. Nevertheless, for $GaN$ both masses are almost the same $[15]$. Therefore, 
the result expressed by Eq. (7)  is valid at this level of approximation. 
 If only the LO-phonons were active, the polaron 
effective mass correction is $Mr(\pi/2)\simeq Mr(0)=7.52\times 10^{-2}$. By contrary for
TO-phonons $Mr(\pi/2)=3.83\times 10^{-5}$ and $Mr(0)=5.61\times 10^{-5}$, showing 
that the interaction with TO-like phonons is responsible for the introduction
of a strong $\eta$-dependence. However, the relative contribution of this last correction
makes the total result for the anisotropy to be small. 

The magnitude of the polaron mass anisotropy is also 
related to some other material factors like the phonon frequencies
and dielectric constant. Other III-V wurtzite semiconductors
like $AlN$ and $InN$ with similar ionicities show roughly the same results$[16]$. By contrast
wurtzite II-VI semiconductors which have larger ionicities but similar phonon frequencies, 
should present a larger polaron mass anisotropy, which could be calculated in the same
way, but experimental Fr\"ohlich constants for those materials are scarce.
It can be expected that
for the case of different wurtzite materials, a higher influence of the electron-phonon
interaction in the effective mass anisotropy could be appreciable.
 
In the results of Ref. $[6]$ for the scattering rates at $T=300\;K$, it is also 
seen that the contribution coming from the TO-like mode is comparatively small. 
The largest value of the TO-like matrix element is reported to be $7\%$ of the 
LO-like matrix element. This indicates that for finite temperatures, the relative amount of 
the TO-like contribution might increase. To investigate that point, we have 
computed for ${\bf k}=0$ the so-called polaronic damping rate in wurtzite $GaN$ 
from the first order imaginary part of the retarded self-energy of Eqn. (3). In our case -as can be 
seen in the corresponding expression-, only the phonon absorption process is 
non-vanishing because for ${\bf k}$ the emission threshold is never reached even 
for finite temperature. The polaron damping rate is given through the equation:

\[{{1}\over{\tau}_{S}}=\alpha_{S}\,\omega_{S}\left(
{\epsilon^{*}_{z}\epsilon^{*}_{\perp}}\right)^{1/2}\,I_{S}; \]
\be
I_{S}={{2}\over{\omega_{S}^{1/2}}}\,\int^{1}_{0}\frac{f_{S}(x)}{\Omega^{3/2}_{S}(x
)}\csc\! h[\beta\hbar\Omega_{S}(x)] dx,
\ee

with $\beta=1/k_{B}T$. 

Fig. 2 shows the dimensionless quantity $I_{S}$ as a function of the temperature. 
Fig. 2(a) corresponds to the LO-like case, and Fig. 2(b) corresponds to the 
TO-like case. It is inmediately seen that the value of the TO-like contribution to 
the polaron damping
rate is always much smaller than the LO-like one; but its magnitude 
increases in one order of magnitude when going from $T=100\;$K to $T=300\;$K, and 
this could be an important effect for a different wurtzite material, where the 
combined values
 of the parameters give rise to a larger electron-TO-like-phonon matrix element.

In summary, we have reported in this work the polaron binding energy and effective 
mass in first order perturbation theory for wurtzite materials. Numerical results 
for the case of $GaN$ show that the magnitude of the coupling between the 
conduction electrons and the polar
optical phonons is large indeed. Both, the LO-like and TO-like 
oscillation 
modes contribute, even when it is seen that the relative weight of the TO-like 
contribution is much smaller than the one coming from the LO-like phonons; 
but it is possible to expect appreciable contributions from the TO-like phonon 
scattering processes at room temperature. The value of the polaron
binding energy indicates the 
great importance that the carrier-phonon interaction might have, for instance, in 
the study of exciton-related optical processes in wurtzite semiconducting 
materials and heterostructures based on them. It is very well known that these materials are the 
subject of interesting applications in optoelectronics, where III-V nitride-based 
blue-green semiconductor lasers have already been produced. 

Obviously, the electron-phonon coupling is strong enough to allow for 
non-negligible contributions from the second order diagrams in the self-energy 
expansion. According to the perturbation Hamiltonian $[6]$, these second order 
diagrams, with two phonon
lines, occur with three different structures: two LO-like phonon lines, two TO-like 
phonon lines, and one LO-like plus one TO-like phonon lines.  For zero 
temperature, in the evaluation of the second order corrections, it 
is necessary to consider
only the contributions coming from the first of these structures: that of two 
LO-like phonon scattering processes. This, in the case of $GaN$, is justified by 
the numerical value above reported for the polaron binding energy in first order 
approximation (not to mention that the magnitude of the effective mass correction 
is considerably
smaller). Nevertheless, for finite temperature, the contributions of TO-like 
phonons in second order might also become significative, possibly for a different 
wurtzite material. The work of calculating these second order corrections, and 
their inclusion in the study of exciton-related optical absorption is already in progress.

\bigskip
{\bf\Large Acknowledgements}

M.E.M.R. gratefully thanks kind hospitality from Dept. of Physics, University of 
Los Andes. F.J.R. and L.Q. acknowledge support from COLCIENCIAS through project 
1204-05-264-94.

\newpage

{\bf\Large Figure Captions}

{\bf Fig.1}: Relative polaron effective mass for $GaN$ as a function of $\eta$, 
the angle between
the electron wavevector and the $c$ axis of the wurzite crystalline structure. 


{\bf Fig.2}: The dimensionless quantity $I_{S}$ as a function of the temperature 
for $GaN$: (a) LO-like phonon absorption process, and (b) 
TO-like phonon absorption process.


\bigskip

\begin{thebibliography}{99}


\bibitem{} M. A. Khan, J. N. Kuznia, A. R. Bhattarai, and D. T. Olson, Appl. Phys. 
Lett. {\bf 62}, 1786 (1993);
M. A. Khan, M. S. Shur, J. N. Kuznia, Q. Chen, J. Burm, and W. Schaff, Appl. Phys. 
Lett. {\bf 66}, 1083 (1995).

\bibitem{} S. Nakamura, M. Senoh, S. Nagahama, N. Iwasa, T. Yamada, T. Matsushita, 
and H. Kiyoku, Appl. Phys. Lett. {\bf 69}, 1477 (1996); Appl. Phys. Lett. {\bf 
70}, 616 (1997).

\bibitem{} {\it Gallium Nitrides and Related Compounds}, edited by R. D. Dupuis, 
J. A. Edmond, F. Ponce, and S. Nakamura, {\it MRS Symposia Proceedings}, {\bf No. 
395} (Materials Research Society, Pittsburgh 1996).

\bibitem{} {\it III-V Nitrides}, edited by F. A. Ponce, T. D. Moustakas, I. Asaki, 
and B. A. Monemar, {\it MRS Symposia Proceedings}, {\bf No. 449} (Materials 
Research Society, Pittsburgh 1997).
 
\bibitem{} F. Bernardini, V. Fiorentini, and D. Vanderbilt, Phys. Rev. B {\bf 56}, 
R10024 (1997).

\bibitem{} B. C. Lee, K. W. Kim, M. Dutta, and M. A. Stroscio, Phys. Rev. B {\bf 
56}, 997 (1997).

\bibitem{} F. Comas, R. P\'erez-Alvarez, C. Trallero-Giner, and M. Cardona, 
Superlatt. and Microstructures {\bf 14}, 95 (1993).

\bibitem{} F. Comas, C. Trallero-Giner, and M. Cardona, Phys. Rev. B {\bf 56}, 
4115 (1997).

\bibitem{} M. E. Mora-Ramos, and D. A. Contreras-Solorio, Physica B {\bf 253}, 
325 (1998). 

\bibitem{} L. Wendler, Physica Status Solidi (b) {\bf 129}, 513 (1985). 

\bibitem{} A. Anselm, {\it "Introduction to Semiconductor Theory".} 
(Prentice-Hall, Englewood Cliffs, NJ, 1981).

\bibitem{} T. Azuhata, T. Sota, K. Suzuki, and S. Nakamura, J. Phys. Condens. 
Matter {\bf 7}, L129 (1995).

\bibitem{} M. Drechsler, D. M. Hofmann, B. K. Meyer, T. Detchprohm, H. Amano, and 
I. Akasaki, Jpn. J. Appl. Phys. 1 {\bf 34}, L1178 (1995).

\bibitem{} G. D. Mahan, {\it "Many Particle Physics".} 2nd. Ed. 
(Plenum, New York, 1990).


\bibitem{} Ch.-H. Kim and B.-H. Han, Sol. Stat. Commun. {\bf 106}, 127 (1998).

\bibitem{} M. E. Mora-Ramos, F. J. Rodr\'{\i}guez, and L. Quiroga (to be published).





\end{thebibliography}
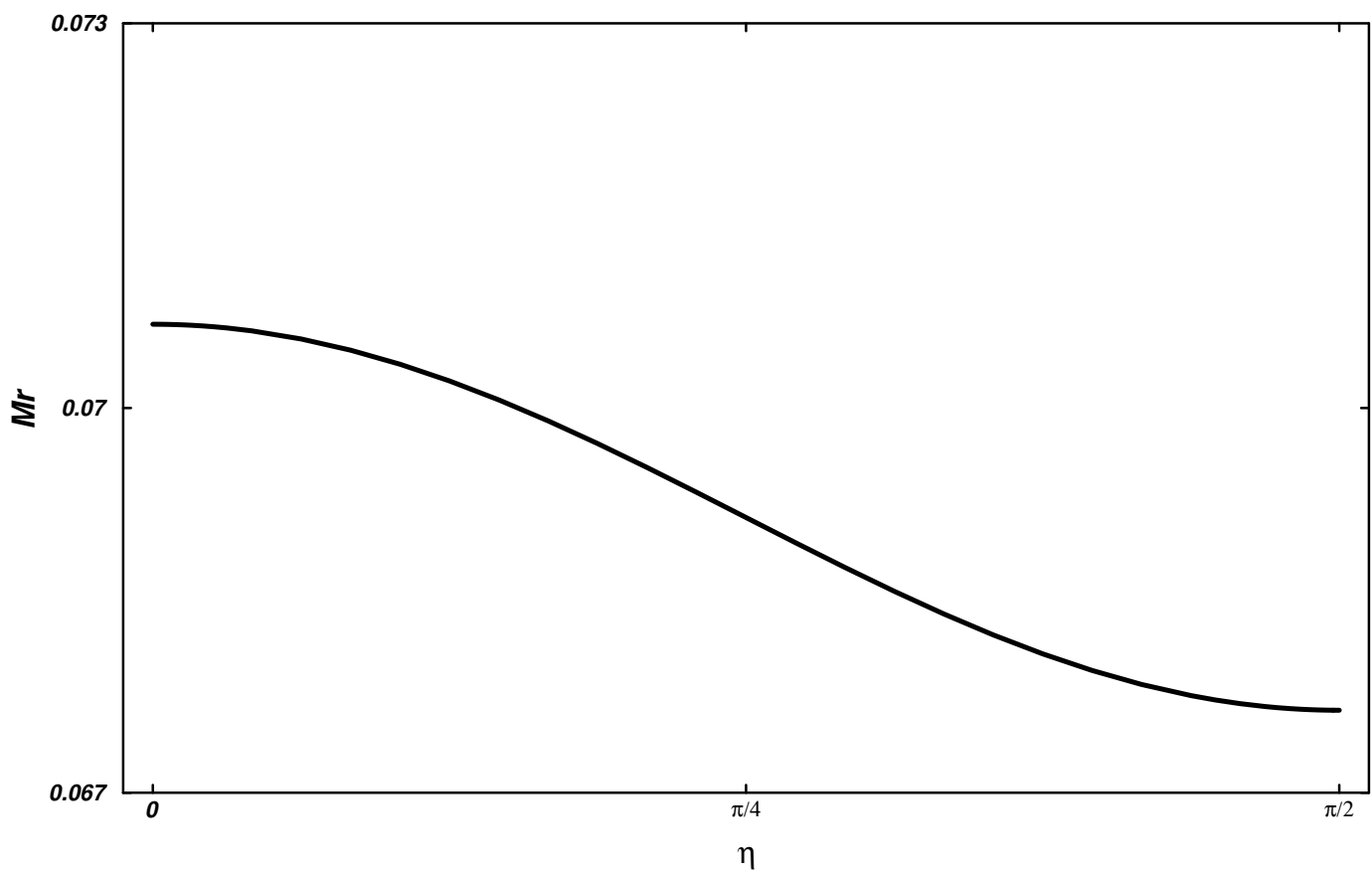
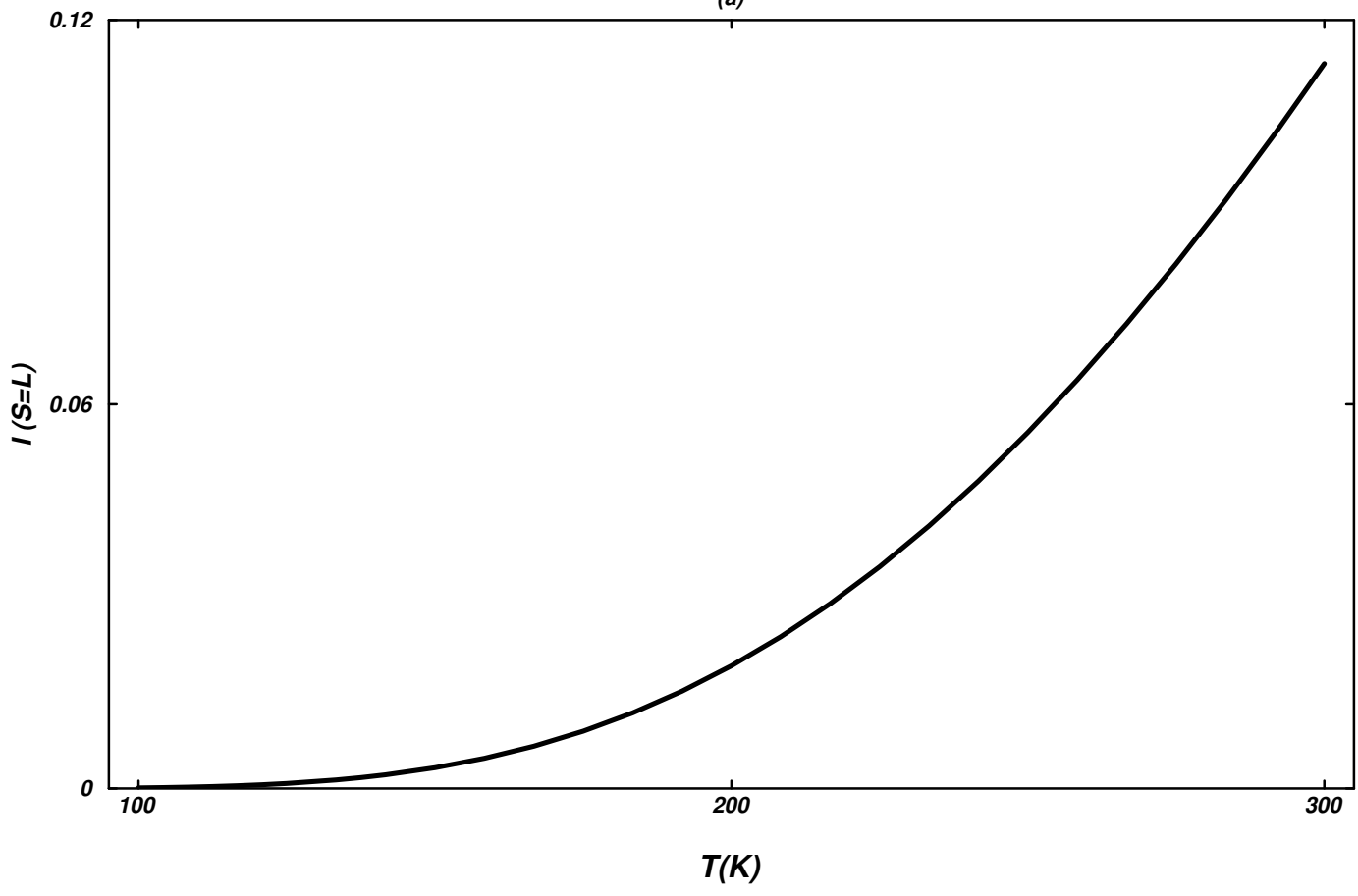
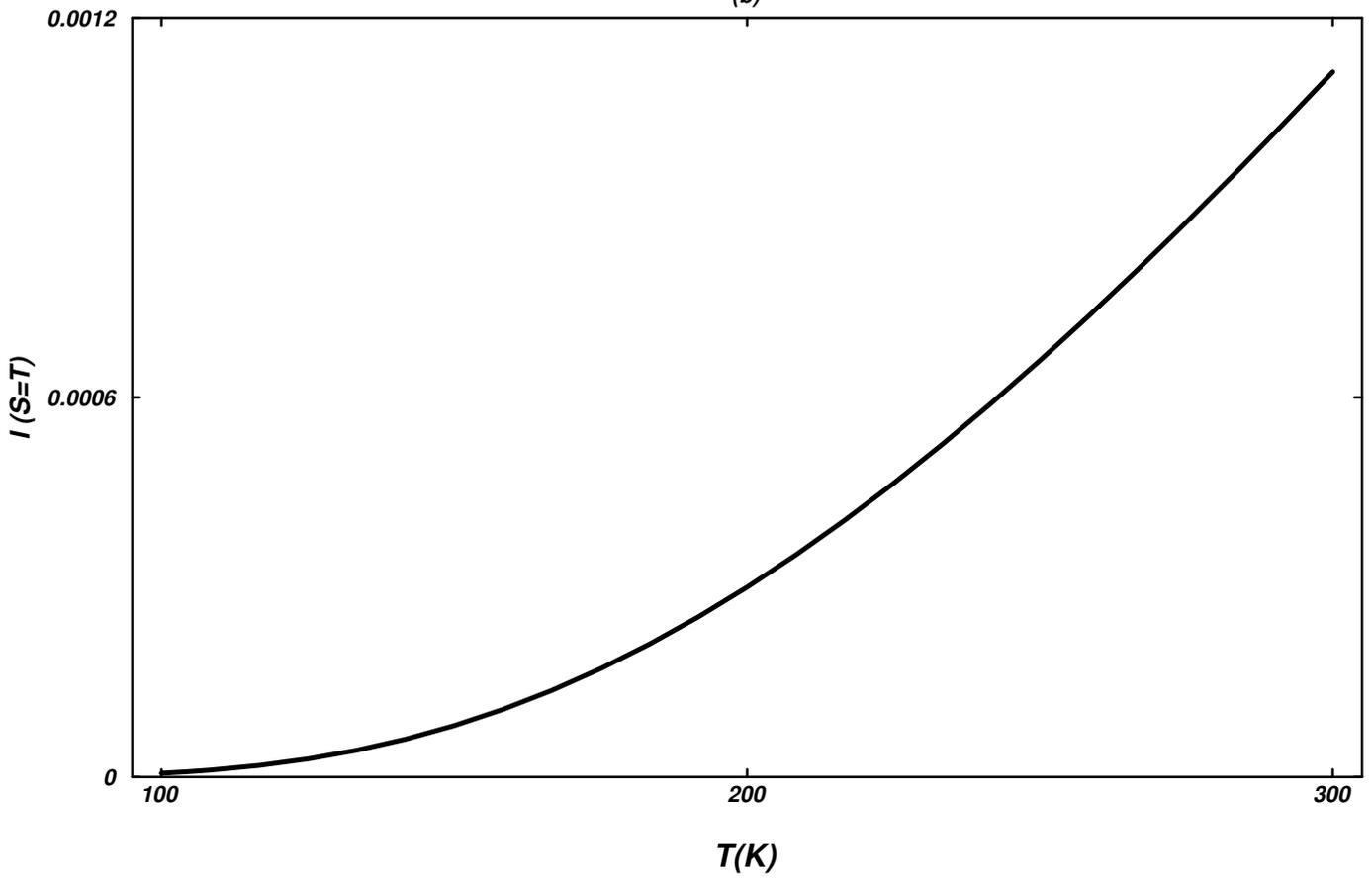
\end{document}